\documentclass[%
 reprint,
 amsmath,amssymb,
 aps,
floatfix,
]{revtex4-1}

\usepackage{graphicx}% Include figure files
\usepackage{dcolumn}% Align table columns on decimal point
\usepackage{bm}% bold math
\mathchardef\mhyphen="2D
\begin{document}

\preprint{APS/123-QED}

\title{3D photonics for ultra-low energy, high bandwidth-density chip data links}% Has to be 75 characters including spaces

%\title{Massively scalable silicon photonic transmitter driven \\ by an integrated Kerr frequency comb}

\author{Stuart Daudlin\textsuperscript{1}, Anthony Rizzo\textsuperscript{1,2}, Sunwoo Lee\textsuperscript{3}, Devesh Khilwani\textsuperscript{3}, Christine Ou\textsuperscript{3}, Songli Wang\textsuperscript{1}, Asher Novick\textsuperscript{1}, Vignesh Gopal\textsuperscript{1}, Michael Cullen\textsuperscript{1}, Robert Parsons\textsuperscript{1}, Alyosha Molnar\textsuperscript{3}, and Keren Bergman\textsuperscript{1,*}}%

\address{\textsuperscript{1}Department of Electrical Engineering, Columbia University, New York, NY 10027 \\
\textsuperscript{2}Air Force Research Laboratory Information Directorate, Rome, NY 13441 \\
\textsuperscript{3} Department of Electrical and Computer Engineering, Cornell University, Ithaca, NY 14853}

%\begin{abstract}
%    Artificial intelligence (AI) hardware is positioned to achieve remarkable computational achievements due to its advanced computer chips working in vast networks that scale to the thousands. However, AI chip development has far outpaced that of the networks that connect them, as chip speeds have accelerated a thousandfold faster than communication speeds over the last two decades. This gap is the largest barrier for scaling AI performance, and results from the disproportionately high energy expended to transmit data, which is two orders of magnitude more intensive than computing. Here we show a leveling of this long-standing discrepancy and achieve the lowest energy and largest scale data link to date, using many channels of light rather than electrical wires. Previously, efforts for large and low-power photonic links were limited to 8 data channels; here we show 80 data channels from one photonic chip and a twofold improvement of the lowest energies consumed in prior works (at 120 femtojoules per bit communicated). To realize these transformative efficiencies, we bond the photonic chip with an electronic chip using a dense three-dimensional technique and double the previously highest density of data transfer (reaching 5.3 tera-bits per second per mm\textsuperscript{2}).  This work represents significant progress in the field of photonics, and will enable further expansion of AI through ultra-energy efficient data communication between every computer in AI networks.
%\end{abstract}

% \date{\today}

\maketitle

\textbf{Artificial intelligence (AI) hardware is positioned to unlock revolutionary computational abilities across diverse fields ranging from fundamental science \cite{carrasquilla2017machine} to medicine \cite{he2019practical} and environmental science \cite{kaack2022aligning} by leveraging advanced semiconductor chips interconnected in vast distributed networks. However, AI chip development has far outpaced that of the networks that connect them, as chip computation speeds have accelerated a thousandfold faster than communication bandwidth over the last two decades \cite{dally2021evolution, mirabbasi2023through}. This gap is the largest barrier for scaling AI performance \cite{wang2023zero++, pati2023computation} and results from the disproportionately high energy expended to transmit data \cite{lee2022beyond}, which is two orders of magnitude more intensive than computing \cite{miller2017attojoule}. Here, we show a leveling of this long-standing discrepancy and achieve the lowest energy optical data link to date through dense 3D integration of photonic and electronic chips. At 120 fJ of consumed energy per communicated bit and 5.3 Tb/s bandwidth per square millimeter of chip area, our platform simultaneously achieves a twofold improvement in both energy consumption and bandwidth density relative to prior demonstrations \cite{rakowski2018hybrid, samanta2023direct}. These improvements are realized through employing massively parallel 80 channel microresonator-based transmitter and receiver arrays operating at 10 Gb/s per channel, occupying a combined chip footprint of only 0.32 mm\textsuperscript{2}. Furthermore, commercial complementary metal-oxide-semiconductor (CMOS) foundries fabricate both the electronic and photonic chips on 300 mm wafers, providing a clear avenue to volume scaling. Through these demonstrated ultra-energy efficient, high bandwidth data communication links, this work eliminates the bandwidth bottleneck between spatially distanced compute nodes and will enable a fundamentally new scale of future AI computing hardware without constraints on data locality.}

%Through eliminating the bandwidth bottleneck between computational nodes and lifting the constraint on data locality in distributed computing systems, this work will enable unprecedented scaling of AI systems with ultra-energy efficient high bandwidth data communication between spatially distanced compute nodes.}

Light, as a medium for communication, has the unique ability to transmit volumes of data with minimal energy loss. This capability not only sparked the revolution of internet-based communication over fibre optic networks, but also holds the potential to significantly expand computing power beyond current capabilities. Specifically, artificial intelligence (AI) is poised to dramatically transform the computational landscape if provided with more efficient data communication between nodes in computer networks \cite{wang2023zero++, wu2023peta}. A critical bottleneck to the full implementation of light-based communication is the conversion of electrical data from inside a computer chip to optical data. At present, data is stored densely in these semiconductor chips in compute nodes, but is sent out of the chip through centimeter-long electrical wires before finally interfacing with optical transmitters in the form of pluggable optical transceivers. The design of these electrical channels results in slow data rates that are not scalable without accounting for a substantial amount of energy consumption \cite{lee2022beyond}. To overcome this bottleneck, electrical channels must be condensed and converted into optical signals within a compact area.

\begin{figure*}
    \centering
    \includegraphics[scale=1.1]{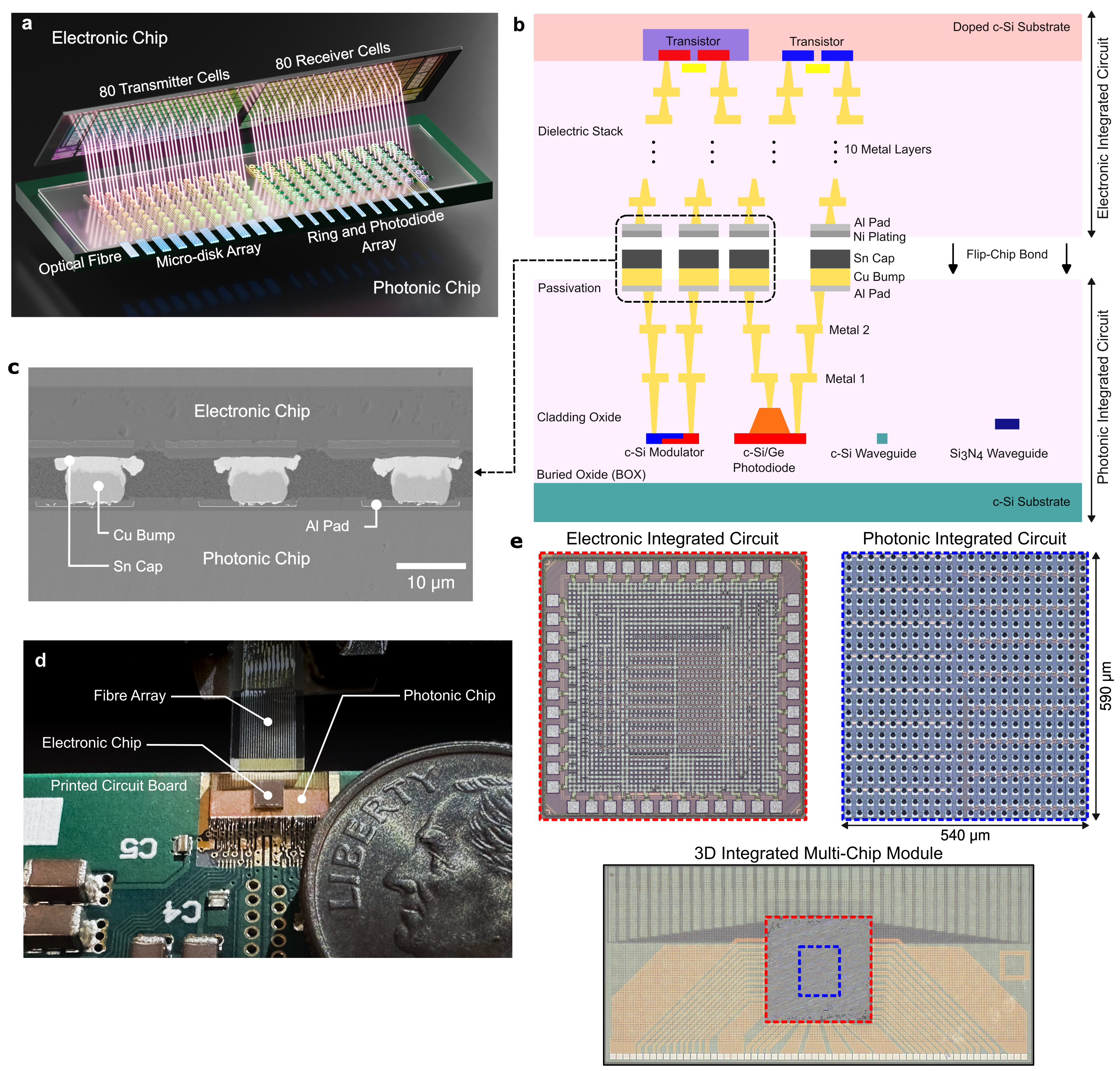}
    \caption{\textbf{3D integrated photonic-electronic transceiver. a,} Illustration of the 3D integrated photonic-electronic system combining arrays of electronic transmitter and receiver cells with arrays of photonic devices. \textbf{b,} Cross-sectional diagram of the electronic and photonic chips and their associated material stacks. Both chips consist of a crystalline silicon (c-Si) substrate, doped silicon devices, and metal interconnection layers. \textbf{c,} A scanning electron microscope (SEM) image of the cross section of the flip-chip bonded electronic and photonic chips. \textbf{d,} Image of the wire-bonded transceiver die-bonded to a printed circuit board and optically coupled to a fibre array with a U.S. dime for scale. \textbf{e,} Microscope images of the standalone and 3D integrated chips with the larger photonic chip beneath the flipped electronic chip. The active photonic circuits occupy the area outlined in blue, while the rest of the photonic chip area is used to fan out the optical/electrical lanes for fibre coupling and wire bonding.}
\end{figure*}

Previously, intensive efforts have produced chip-scale transmitters and receivers (transceivers) towards this goal but have been marked by a lack of efficiency or scale. These works build on the field of integrated photonics, a technology that aggregates a multitude of optical components onto a single integrated chip. In particular, silicon is highly appealing as a material platform for integrated photonics since it can leverage the tremendous investment in the complementary metal-oxide-semiconductor (CMOS) infrastructure used to fabricate microelectronics chips \cite{hochberg2010towards}. The silicon photonics technology platform includes devices such as micro-resonator-based modulators \cite{xu2005micrometre, timurdogan2014ultralow}, filters, and germanium photodiodes \cite{michel2010high} that are compact, efficient in their electrical-to-optical and optical-to-electrical conversions, and scalable to many wavelength channels \cite{rizzo2023massively, rizzo2022petabit}. To date, the largest of these systems is comprised of 64 channels of photonics and electronics on a single chip and achieves 240 femtojoules per communicated bit (fJ/bit) by the transmitter \cite{wade2018bandwidth, sun2020teraphy, wade2021error}. However, this system has receiver energy consumption above 1000 fJ/bit and has a limited density from the lateral arrangement of photonics and electronics on the same two-dimensional chip. While this monolithic integration of CMOS transistors alongside photonic devices on the same chip may appear highly appealing  \cite{sun2015single, atabaki2018integrating}, this configuration `freezes' transistors at a given node size and thus cannot benefit from the further energy efficiency, size, and speed gains of moving to more advanced CMOS nodes. Alternatively, three-dimensional (3D) integration combines a more efficient, leading edge CMOS node electronic chip and a separate photonic chip to improve on these limitations. Ongoing 3D efforts have demonstrated sub-200 fJ/bit powers from transmitters \cite{zheng2011ultralow, rakowski2018hybrid, samanta2023direct, ban2023highly} and receivers \cite{rakowski2018hybrid, samanta2023direct, saeedi201625}, but the chip-to-chip bond spacings are either significantly larger than the devices themselves \cite{zheng2011ultralow, rakowski2018hybrid, ban2023highly, saeedi201625} or rely on emerging hybrid bonding technology \cite{samanta2023direct}. Furthermore, 3D integrated transceivers have not scaled to more than eight channels \cite{zheng2011ultralow} and have yet to achieve both transmitter and receiver powers below 100 fJ/bit. 

Here, we present the most energy-efficient conversion between electrical data and optical data and the highest density of data transmission from an integrated chip-scale system. Our novel approach to photonic transceivers simultaneously addresses energy efficiency and bandwidth density scaling for future computing systems. This transceiver demonstration is scaled up to 80 channels while having a scaled-down energy consumption through low-capacitance connections between low-capacitance photonics and co-designed CMOS electronic circuits. The data signaling rate per channel is relatively low at 10 gigabits per second (Gb/s) per channel, which permits the receiver electronics to operate in an ideal regime for minimized energy consumption (Methods). The large array of channels compensates for the low per channel data rates and delivers a high aggregate data rate of 800 Gb/s in a compact area of only 0.32 mm\textsuperscript{2}. From the perspective of interfacing to a processor, a large array of low data rate channels relaxes signal processing and time multiplexing of the low data rate streams native to the processor \cite{miller2017attojoule, georgas2011addressing, zhu2020photonic}. Furthermore, wavelength-division multiplexing (WDM) sources for these numerous data streams are becoming readily available with the advent of chip-scale micro-combs \cite{gaeta2019photonic, rizzo2023massively}. Our demonstration unlocks the tremendous potential of light as a high bandwidth and energy-efficient inter-chip communication medium, offering an immediate solution to the pressing challenge of AI scaling.

\section*{Results}

\begin{figure*}[h!]
    \centering
    \includegraphics[scale=0.9]{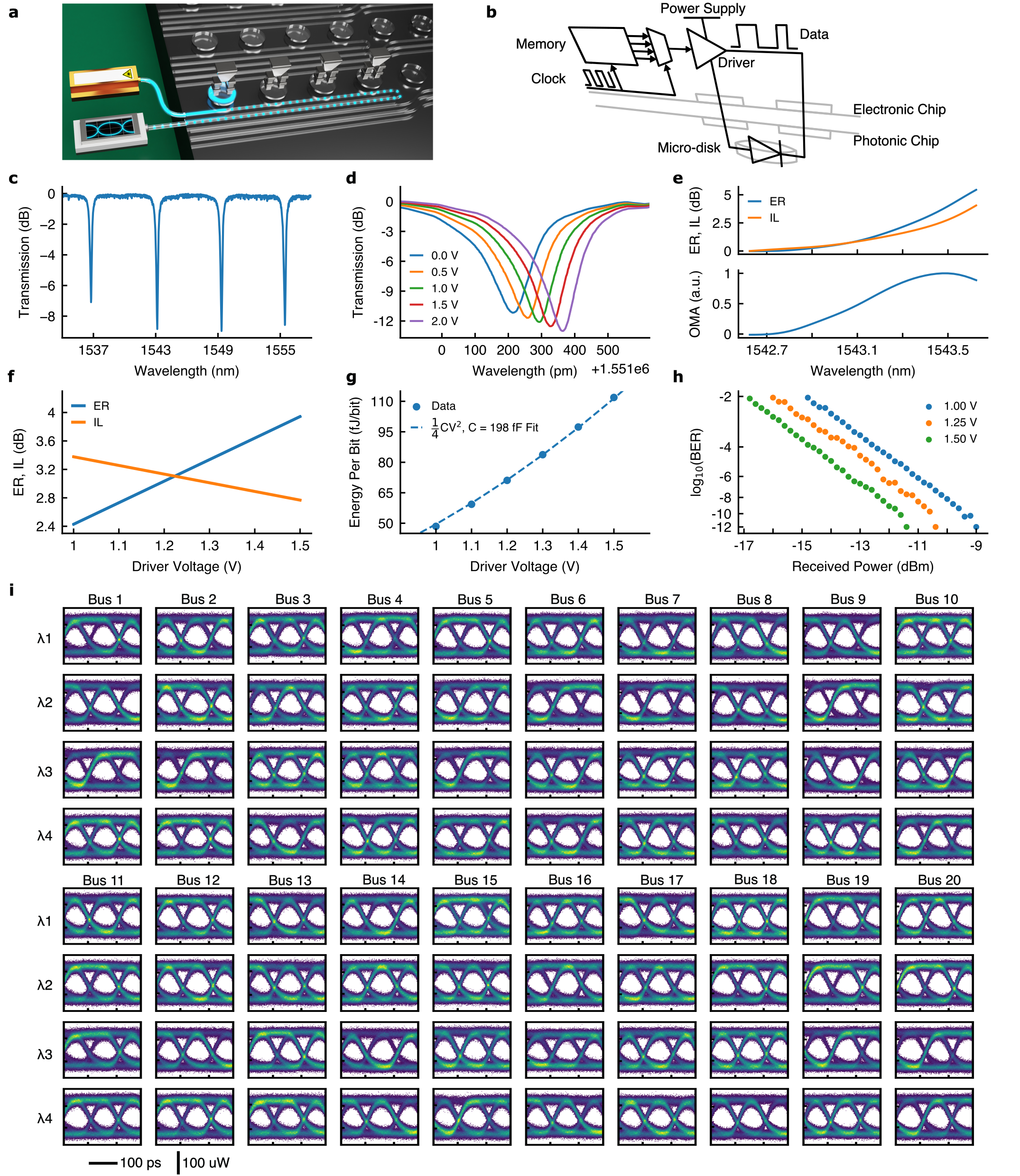}
    \caption{\textbf{Transmitter characterization and performance. a,} Illustration of the transmitter experimental test setup showing a single wavelength laser channel modulated by the transmitter and measured on an oscilloscope. \textbf{b,} Transmitter cell circuit schematic. \textbf{c,} Optical spectrum of the four channel transmitter bus. \textbf{d,} Resonance shift of a representative micro-disk as a function of reverse-bias voltage applied to the vertical p-n junction. \textbf{e,} Transmitted signal extinction ratio (ER, power of ‘1’ bit divided by power of ‘0’ bit), insertion loss (IL, power of ‘1’ bit divided by power before the modulator), and normalized optical modulation amplitude (OMA, normalized power of ‘1’ bit minus power of ‘0’ bit) with a 1.5 V driver voltage. \textbf{f,} ER and IL at maximum OMA for a range of driver voltages. \textbf{g,} Measured energy consumption of the transmitter array for a range of driver voltages and a $\frac{1}{4}$CV\textsuperscript{2} fit, where C is the capacitance charged by the driver voltage V. \textbf{h,} Bit error ratio (BER) measurement of the modulated signal input to a commercial receiver at 1 V, 1.25 V, and 1.5 V driver voltages; received power is the average signal power at the commercial receiver. \textbf{i,} Eye diagrams for all 80 modulators on the photonic chip at 10 Gb/s/modulator and 1 dBm input laser power.}
\end{figure*}

We implement this high-density transceiver using compact photonic devices and dense, 28-nm node co-designed electronic circuits; however, the total density ultimately depends on the 3D bond spacings. To address this, we employ a high-density bonding process using copper pillar bumps. An electroplating process is used to form bumps on the photonic chip with copper pedestals capped with a layer of tin. The copper-tin bumps are then bonded to a nickel-plated electronic chip under a thermal and compressive force bringing the chips together. Figure 1b illustrates the layers on the electronic chip, photonic chip, and the bonding metals. We push the limits of this bonding technology by using a 15 $\mu$m spacing and 10 $\mu$m bump diameters (25 $\mu$m pitch) in an array of 2,304 bonds. This process balances two potential failure modes for such close spacing: excessive tin causing flow and electrical shorting to adjacent bonds during bonding, and insufficient tin leading to brittle bonds \cite{li2021scaling}. We test our bonding process using cross-sectional scanning electron micrographs of the bonds, shown in Figure 1c, and by measuring the force needed to separate the bonded chips. The cross-section analysis reveals that tin does not flow to adjacent bonds, while the shear test demonstrates a robust 2.1 kg (114.9 MPa) force required to separate the bonded dies. Modeling and measurements show each pair of bonds (for a signal and ground) has a 10 fF capacitance (Methods). Figure 1d shows the assembled transceiver wire-bonded to a printed circuit board and optically coupled to a fibre array (Methods), while Figure 1e shows microscope images of the face-up electronic and photonic chips, and the face-down electronic chip flip-chip bonded on top of the larger photonic chip. This bonding technique provides an ideal platform to achieve the required density for chip-to-chip data communication links.

The 3D integrated chip contains an array of 80 transmitter cells and 80 receiver cells; these cells are organized into 20 waveguide buses with four wavelength channels per bus. Each transmitter cell has a local memory in the electronic chip that stores a pseudo-random bit sequence. A periodic clock signal triggers the electronics, and the transmitter cell electronics send out the programmed bit sequences as voltage pulses incident on the photonic modulator electrodes. These voltage pulses blue-shift the micro-disk resonance from a blocking to a non-blocking state and thus modulate an on-resonance laser line. Figure 2a illustrates the transmitter experiment and Figure 2b shows a schematic of the transmitter cell, while Figure 2c shows the spectrum of the modulator bus with four micro-disk resonances. After the transmitter characterization, we test the receivers, which function similarly: in each receiver cell, wavelength channels carry signals on the photonic chip and micro-rings selectively drop wavelengths onto each respective photodiode. The electronic chip then amplifies the photocurrent generated by the photodiode and writes the data into the local memory of each receiver cell, as illustrated in Figure 3a. Figure 3b shows a schematic of the receiver cell. For performance characterization, an on-chip circuit compares the receiver memory to expected data and keeps an error count that is periodically read out of the chip. This architecture of transmitters and receivers fills the array of channels in the dense area provided by the bonding process.

\begin{figure*}
    \centering
    \includegraphics[scale=1]{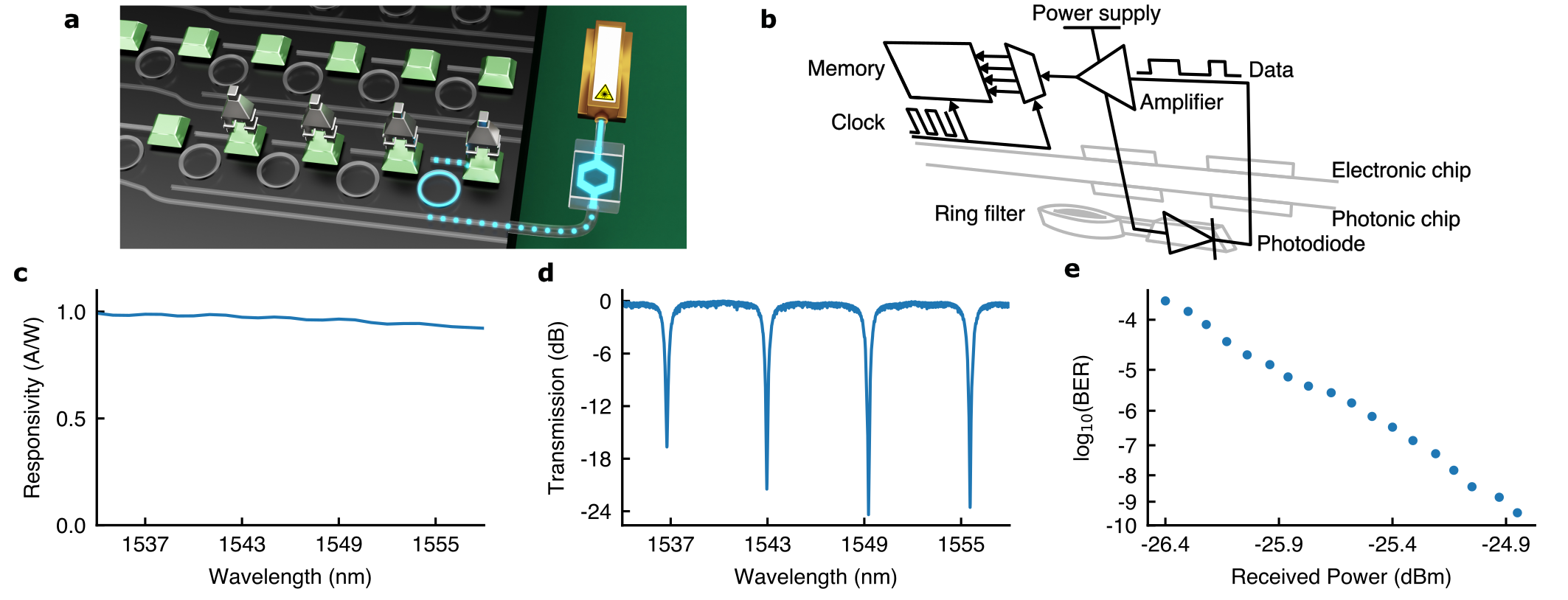}
    \caption{\textbf{Receiver characterization and performance. a,} Illustration of the receiver test setup showing a laser line modulated by a commercial transmitter and received by the 3D integrated photonic-electronic receiver. \textbf{b,} Receiver cell circuit schematic. \textbf{c,} Responsivity (light to electrical current conversion efficiency) measurement of the photodiode. \textbf{d,} Optical spectrum of receiver bus. \textbf{e,} Bit error ratio (BER) test of a receiver cell using a commercial transmitter signal; received power is the average signal power at the photodiode.}
\end{figure*}

The transmitter cell within the 80-cell array consumes 50 fJ/bit when driving the micro-disks with a 1 V swing. This power is dynamic, equal to $\frac{1}{4} \textrm{CV}^{2}$, where C is the capacitance being charged or discharged during a bit transition, and V is the charged-to-discharged voltage \cite{miller2012energy}. The vertical p-n junction micro-disk enables a low voltage drive by featuring a higher overlap of the p-n depletion region and the optical whispering gallery mode of the disk compared to lateral junctions \cite{timurdogan2014ultralow}, and results in an electrical-to-optical response of 75 pm resonance shift per applied volt (Fig. 2d). We further characterize this response with the dynamic insertion loss (IL, the power of a ‘1’ bit divided by the power before the modulator) and extinction ratio (ER, the power of a ‘1’ bit divided by the power of a ‘0’ bit). Figure 2e shows these metrics, captured from the modulated signal output of a transmitter cell driven at 1.5 V. In this measurement, the laser wavelength moves into the shifting resonance and the optical modulation amplitude (OMA, the power of ‘1’ bit minus the power of ‘0’ bit) of the output signal increases, reaching a maximum at 2.5 dB IL and 4 dB ER. Figure 2f shows ER and IL at maximum OMA for driver voltages between 1 and 1.5V. These high ERs and low ILs per volt enable a reduced V in $\frac{1}{4} \textrm{CV}^{2}$. Capacitance sources include the micro-disk p-n junction (128 fF), bond pads (10 fF), and capacitances within the driver circuit (61 fF), combining for a total expected capacitance of 199 fF (Methods). These capacitances exhibit low values through micro-disk compactness, miniaturized bonds, and careful design in the 28 nm electronic chip technology. Figure 2g shows the transmitter energy consumption as all 80 modulators are transmitting data with drive voltages ranging between 1 V to 1.5 V. The $\frac{1}{4} \textrm{CV}^{2}$ model fits 198 fF total capacitance per cell using this data, aligning closely with the expected 199 fF from the independently measured and modeled devices. Next, we record eye diagrams for each of the 80 transmitters on the chip with a drive at 10 Gb/s/transmitter and 1 dBm laser power before the modulator (Fig. 2i). As every transmitter modulates, we measure current and voltage on the transmitter power supply for the previous energy consumption (Fig. 2g). With no optical amplification, the oscilloscope receiver is the limiting factor of the eye qualities with an 8 uW input-referred root mean square noise (denoted as input-referred noise here on). All 80 eye diagrams in the array are open and uniform, which confirms the high-yield of the bonding process and validates our many-channel approach. As a further confirmation of transmitter signal quality, a bit error ratio (BER) test with a reference receiver demonstrates error-free performance (BER $<$ 10\textsuperscript{-12}) down to a receiver noise-determined power for 1 V, 1.25 V, and 1.5 V modulator drives (Fig. 2h). In the following section, our on-chip receiver shows a dramatically reduced input-referred noise of 480 nW. The 80 channel photonic transmitter array outputs an aggregate data rate of 800 Gb/s and occupies an area of 0.15 mm\textsuperscript{2}, demonstrating an unprecedented bandwidth density of 5.3 Tb/s/mm\textsuperscript{2}.

The receiver cell consumes 70 fJ/bit when receiving a 10 Gb/s signal at -24.85 dBm average power with a $4 \times 10^{-10}$ BER. The receiver spends energy as a static biasing of the electronic amplifier. The photodiode is a vertical p-silicon, i-germanium, n-germanium diode that efficiently converts optical signals to electrical current with an efficiency of 1 A/W (Fig. 3c). The capacitance of this photodiode is crucial since receiver noise is proportional to the amplifier input capacitance and is compensated by static biasing power (Methods). Minimizing this noise is critical for reducing laser power sourced into the link and improving energy efficiency. With a measured photodiode capacitance of 17 fF and pad capacitances of 10 fF, the simulated input-referred noise is 300 nW. A BER test is used to evaluate this performance; a signal from an ideal modulator is used to send a 10 Gb/s data stream into the chip from which on-chip circuits measure errors in the received bits. On the photonic chip, a ring resonator on a four-channel bus filters the modulated signal to a photodiode. Figure 3d shows the four-channel bus spectrum. We then gradually reduce the signal power while counting errors on the electronic chip, obtaining the BER curve in Figure 3e. This test reveals that the receiver has a sensitivity of -24.85 dBm for a $4 \times 10^{-10}$ BER, resulting in a measured input-referred noise of 480 nW using the 19 dB ER signal. We record the power of the receiver from its power supply at -24.85 dBm input optical power with 70 fJ/bit consumption. This result, along with the transmitter cell performance, demonstrates that both the receiver and transmitter consume less than 100 fJ/bit while supporting a massive bandwidth transfer of 800 Gb/s from the dense array.

\begin{figure*}
    \centering
    \includegraphics[scale=1]{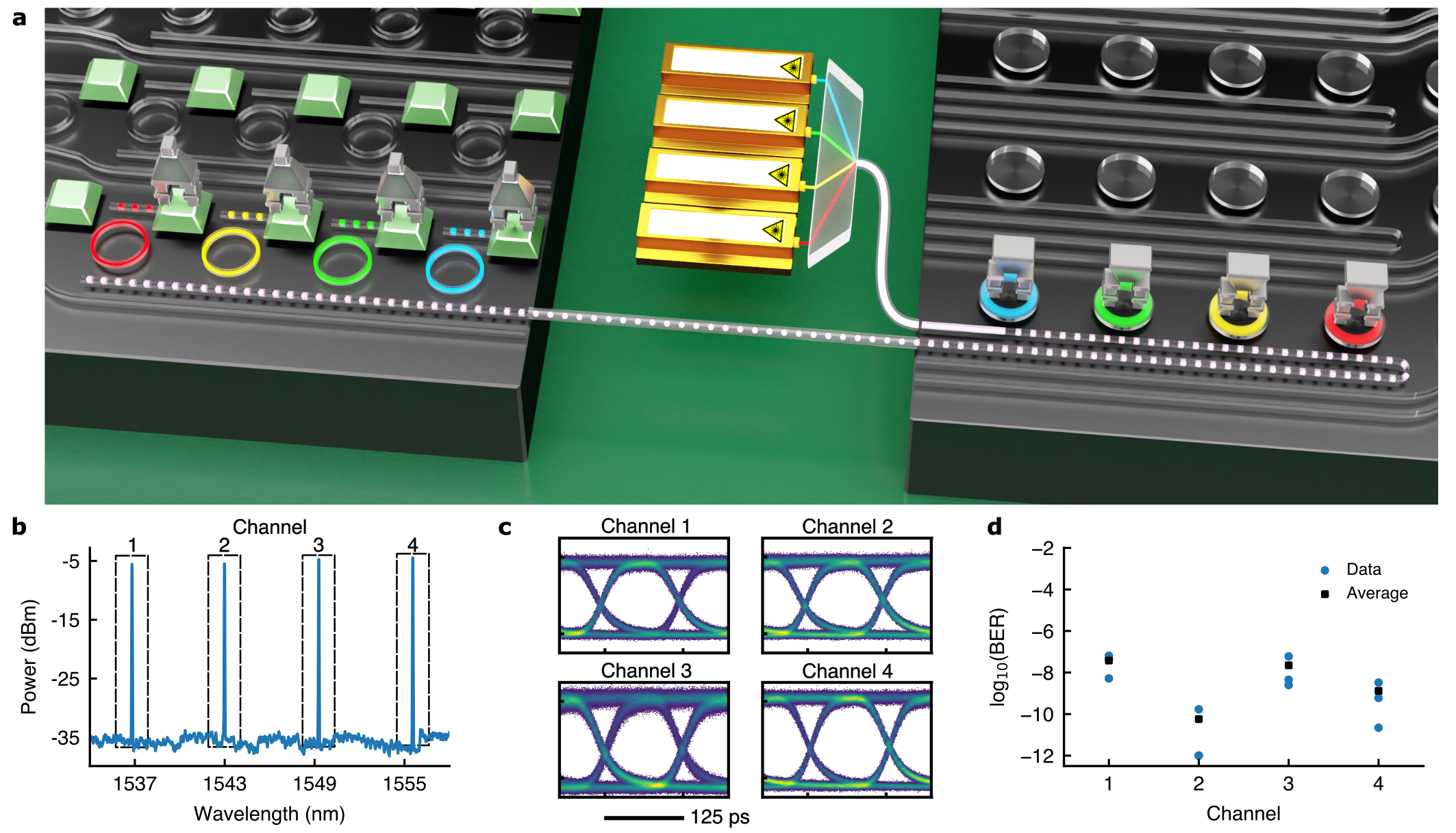}
    \caption{\textbf{Transmitter to receiver data communication link demonstration. a,} Illustration of the transmitter to receiver link showing a 3D integrated photonic-electronic transmitter modulating four laser lines and a separate photonic-electronic receiver converting the four data channels back to the electrical domain. \textbf{b,} Spectrum of the link laser source. \textbf{c,} Eye diagrams of the four channels after the transmitter. \textbf{d,} Bit error ratios (BERs) of the data channels after the receiver.}
\end{figure*}

The transmitter and receiver cells independently demonstrate a combined 120 fJ/bit; we next connect them and validate their combined performance. Optical fibre connects two separate transceivers as a complete data communication link, with one transceiver functioning as a transmitter and the other as a receiver (Fig. 4a). A shared clock synchronizes the two electronic chips, and programmable clock delays in each receiver cell align the transmitted data with the receiver sampling point. A laser diode array provides four wavelength channels at -5 dBm power per channel, which feed into a bus on the transmitter chip. Figure 4b shows the spectrum of these laser channels. Individual transmitter cells simultaneously modulate each wavelength at 8 Gb/s at a 1.5 V drive, resulting in open eye diagrams for each channel (Fig. 4c). The signal powers are too low for detection by the diagnostic oscilloscope receiver, so we amplify the signal before the oscilloscope and normalize the eye diagrams. However, the signals do not require amplification for the electronic-photonic receiver due to its high sensitivity. The average power per channel at each receiver photodiode is -19.5 dBm. On chip error counters record errors in each channel in one-minute intervals, revealing a maximum recorded BER of $6 \times 10^{-8}$ and a minimum count of no errors in the interval, denoted as 10\textsuperscript{-12}, in Figure 4d. This result shows that the transmitters and receivers within the 3D integration can form a complete low-power, high-bandwidth link needed for next-generation computing systems.

\section*{Discussion}

Integrated photonic chips present a promising low-power platform to address the data transfer demands of AI computing. Here, we realized this promise by demonstrating a scaled-up array of 80 channels on a single electronic-photonic, densely 3D integrated transceiver. This multi-chip module consumes minimal energy by virtue of the large number of channels, cutting-edge low-capacitance bonding technology, co-designed electronic-photonic circuits, and advanced devices used. While our demonstrated system achieves record performance in terms of energy efficiency and bandwidth density, the performance can be further improved in future implementations. Although the micro-disks used in this demonstration exhibit high performance, resonant modulators can be developed with lower capacitance \cite{timurdogan2014ultralow, sun2018128} and a higher electro-optical response \cite{timurdogan2014ultralow, gevorgyan2022miniature}, both of which would decrease the dynamic power of the transmitter. Similarly, on the receiver side, lower capacitance photodiodes \cite{lischke2015high, chen20161} could reduce the power and noise of the receiver architecture. However, miniaturizing photodiodes requires considering a loss of responsivity \cite{chrostowski2015silicon}, presenting complex link-level tradeoffs. Additionally, the energy consumption of the electronic circuits can be further reduced by moving to a more advanced CMOS node. While our demonstrated bonding technology is approaching the limit of how closely spaced tin bonds can be made, further density scaling could be realized through the development of hybrid bonding \cite{samanta2023direct}. However, after achieving the low capacitance value of the bonds demonstrated in this work, pursuing a further reduction in bond capacitance would yield diminishing returns in terms of energy efficiency. 

Reduced chip-to-fibre optical losses can improve the loss budget of our demonstrated link, and laser powers may be as low as 47 fJ/bit with a distributed feedback laser and 30 fJ/bit for a scalable, high channel count comb laser (Supplementary Note 1). Furthermore, silicon resonators are sensitive to temperature and fabrication variation, requiring thermal control circuits \cite{sun201645, hattink2022streamlined} and, for minimized power, reduced heat leakage into the environment using methods such as a silicon substrate removal around resonators \cite{ban2023highly, masood2013comparison, rizzo2023ultra}. Detailed wafer-scale resonance variation data and approximate thermal energy contributions across a range of scenarios provide a best case thermal energy consumption of 71 fJ/bit (Supplementary Note 2). Additionally, the photonic circuits are highly polarization-sensitive and require polarization-maintaining fibre or the addition of polarization management circuits \cite{chrostowski2015silicon, ma2015symmetrical, park2019ring}. Finally, while we demonstrate high bandwidth density, a higher per-fibre bandwidth and photonic chip-edge bandwidth density can be achieved with wavelength scaling by cascading more arrayed channels onto fewer waveguide buses \cite{rizzo2023massively, rizzo2022petabit}. This architecture can combine with chip-scale frequency combs to generate hundreds of wavelength channels \cite{rizzo2023massively}.

While the potential impact of this technology is evident for the advancement of energy-efficient AI computing, its use may extend to far reaching applications. These low-power, massively parallel optical links could enable pervasive device connectivity and transform computing by streamlining resource allocation through optically-linked, disaggregated, and reconfigurable computing and memory resources \cite{zhu2020photonic, jouppi2023tpu, gonzalez2022optically, michelogiannakis2022case}, revolutionizing the computing landscape over the next decade and beyond. 

\section*{Methods}
\textbf{Transceiver assembly.} Separate CMOS foundries are used fabricate the electronic and photonic chips. The photonic chips were fabricated through the American Institute for Manufacturing Integrated Photonics (AIM Photonics) on a custom 300 mm silicon-on-insulator wafer. The AIM Photonics process design kit (PDK) includes the micro-disks, ring filters, and photodiodes \cite{fahrenkopf2019aim}. The electronic chips were fabricated through Taiwan Semiconductor Manufacturing Company Limited (TSMC) on a shared multi-project wafer in a 28 nm CMOS process node. Both chips feature a square array of 2,304 aluminium pads at a 25 $\mu$m pitch on their top surfaces. These pads connect to lower metal layers within each chip and the devices on the silicon layer. 

The photonic and electronic chips are then processed post-fabrication before bonding their pad arrays together. In this step, we core the 300 mm photonic wafer to a 200 mm wafer and a wafer-level process is used to bump its pads with electroplated layers of copper and tin. The electronic chips, received as individual 1.6 mm\textsuperscript{2} units from a shared wafer, are unsuitable for wafer-level photolithography-based processes. Alternatively, we adopt a chip-level process of electroless nickel plating, followed by an additional layer of immersion gold plating to prevent nickel oxidation. After dicing the bumped photonic wafer into 6.5 mm by 3 mm chips, a thermo-compression bond is used to connect the bumped photonic chips to the plated electronic chips. 

To power and operate the transceiver, we create electrical connections to the electronic chip through the bonds to the photonic chip. Metal layers on the photonic chip wire these connections to large electrical pads on an exposed edge of the photonic chip. Wire-bonds connect these pads to a printed circuit board (PCB), which connects to: (i) a micro-controller that programs the electronic chip, (ii) power sources that supply the electronic chip voltage rails, and (iii) a radio-frequency (RF) clock generator for the 5 GHz clock of the electronic chip. This clock line is a coplanar RF waveguide on the PCB with a matched impedance to 50 $\Omega$ RF cables. Optical fibres couple light to waveguide buses through silicon nitride edge couplers; these couplers are on the side of the photonic chip that is opposite the wire-bond pads. A micro-positioner is used to align a standard single mode fibre v-groove array with the edge-couplers, which are spaced at a 127 $\mu$m pitch. The assembly procedures of photonic wafer bumping and bonding, electronic chip plating, and wire-bonding are conducted at Micross AIT, CVI, and Cornell University, respectively. \\

\textbf{Capacitance models and measurements.} We identify several sources of capacitance that affect energy efficiency: (i) chip pads, (ii) bump parasitics, (iii) the micro-disk junction, (iv) the photodiode junction, and (v) electronic driver capacitance. Extended Data Figure 1a depicts these capacitance sources. These capacitances are determined through electrostatic simulations, circuit model simulations, and empirical measurements. Focusing initially on the bumps and photonic chip pads, an electrostatic solver (Ansys Maxwell) is used to simulate the photonic chip pad-to-substrate capacitance (4 fF) and bump-to-bump parasitic capacitance ($< 1$ fF). Similarly, we simulate an electronic-chip pad model with extracted parasitic capacitances (Cadence Virtuoso), yielding an electronic chip pad capacitance of 6 fF. This extracted circuit model simulation also results in an effective 61 fF capacitance inside the electronic driver.

Experimental methods are used determine the capacitances of the micro-disk and photodiode junctions, along with a validation of the simulated pad capacitance. A vector-network analyzer (VNA, Keysight P5007A) is used to record an electrical RF reflection from probed devices. We fit the magnitude and phase of the reflected wave across RF frequencies to a reflection from a lumped complex impedance. Extended Data Figure 1b shows the imaginary impedances of the devices and their associated fitted capacitor impedances. During these measurements, an RF bias tee provides a DC reverse bias voltage to the device. An electronic calibration module (Keysight N4693D) is used to calibrate out the VNA response of the RF cable up to the output of the bias tee. The bias tee connects to a 25 $\mu$m pitch RF probe (FormFactor InfinityXT) and this probe response is calibrated out of the results by de-embedding its unlanded response as an electrical open. After this calibration, measurements of the photonic pad and bumped photonic pad show capacitances of 3 fF and 4 fF, respectively. Subsequent measurements are used to first de-embed these pad responses and then measure the photodiode and micro-disk capacitances. The measured photodiode capacitance is 17 fF. 

As anticipated from a p-n junction, the micro-disk capacitance decreases with an increasing reverse bias. Extended Data Figure 1c shows the measured capacitance functions of the four micro-disks. The energy spent per bit transition is the integral of this capacitance function weighted by the difference of supply voltage and output voltage \cite{miller2012energy}. A midpoint Riemann sum of the micro-disk capacitances, weighted by reverse bias voltage subtracted from 1 V, between 0 and 1 V bias voltage, yields an effective disk junction capacitance for dynamic energy consumption of 128 fF (averaged across the four micro-disks). A summation of the pad, driver, and junction capacitances results in a 199 fF transmitter capacitance. This result is in excellent agreement with the capacitance directly measured from the energy consumption of the transmitter (198 fF), validating the capacitance models and measurements. \\

\textbf{Transmitter characterization.} Each transmitter result is experimentally measured from the 3D integrated photonic-electronic chip. An exception is the DC electro-optic response measurement, for which a separate photonic chip with the same modulator design is used. We apply a voltage to a probe to reverse bias the modulator at varying DC voltages, and use an optical spectrum analyzer (Keysight 8164B) to record each response. The remainder of the transmitter, receiver, and link characterizations employ an optical switch (Polatis 1000n 24x10). This switch optically connects equipment and devices-under-test and minimizes fibre mating cycles. This approach streamlines the measurement process and eliminates potential power discrepancies that might stem from fibre mating inconsistencies. An optical spectrum analyzer (Aragon BOSA 400) is used to measure the transmitter bus spectrum. For dynamic data transmission, a micro-controller is used to write a different PRBS6 pattern into the 64-bit registers in each of the 80 transmitter cells. Next, all modulators transmit this data simultaneously as the electronic chip is clocked, and all data registers are driven out of the chip to their respective micro-disks. The 64-bit pattern transmitted by each modulator repeats indefinitely as the chip clock is running. In this state, we record the eye diagrams of each modulator, dynamic characteristics of the modulators, and transmitter energy consumption from the electronic driver array voltage rail. A narrow linewidth tunable laser (Santec TSL-210) is used as the light source in these measurements. Laser light travels through a fibre polarization controller and then into the chip. An oscilloscope (Tektronix DSA8300 with an 80C01 Optical Sampling Module) is used to receive modulated light for dynamic characterization and eye diagrams. In the bit error ratio test, modulated light initially passes through a variable optical attenuator (VOA) before reaching a commercial receiver (Thorlabs RXM40AF). The commercial receiver converts the optical signals into electrical signals that are read by a bit error ratio tester (BERT, Anritsu MU195040A). We sweep the received optical power with the VOA and record errors from the BERT to construct the transmitter BER curves. \\

\textbf{Receiver characterization.} An ideal modulation source and an on-chip bit-error checker circuit are used to characterize the receiver cell performance. Separately, a tunable laser (Keysight 8164B) and a DC electrical probe landed on photodiode pads are used to measure the photodiode responsivity; the probe applies a reverse bias voltage and senses photocurrent from a known input laser power. Next, for the dynamic characterization, we use an ideal modulation source (Thorlabs MX35E) consisting of an internal laser and a lithium niobate Mach-Zehnder modulator. A pulsed pattern generator (PPG, Anritsu MU195020A) is used to drive the modulator with a repeating 64-bit PRBS6 pattern. The signal travels through fibre and a polarization controller before coupling into the photonic chip. Voltage is applied to a doped-silicon resistor adjacent to the ring filter to generate heat and tune the ring filter resonance to the desired wavelength channel. The ring resonator drops the signal to a photodiode, which converts it from light into photocurrent for the electronic chip to then amplify. Timing circuits continuously write the received bits into a 64-bit long memory in a cycle. For timing, a programmable timing offset circuit in the electronic chip and a timing offset of the PPG align the incoming data to the receiver sampling point. A split clock source synchronizes the receiver chip and PPG clock frequencies. As the final step, readouts from the serial programming port display the saved received bits and confirm data reception. However, the serial port cannot update fast enough to give a bit error ratio in a short time frame. Instead, an on-chip error-counter circuit in each cell compares the received memory with pre-programmed expected bits and, if there is a discrepancy, it adds an error to an on-going count. Readouts from the serial port display this count and we obtain a BER curve as we sweep signal power using a VOA inside the ideal modulation source. \\

\textbf{Link demonstration.} The link demonstration combines the previously described experimental setups of the transmitter and receiver. A microcontroller sets the sent bits in a transmitter chip and a second microcontroller reads the received bits in a separate receiver chip. Four channels of data transmit through the link at 8 Gb/s/channel and serial port readouts from the receiver record this data, along with an on-going error count for each channel. A shared clock signal synchronizes the two transceivers and a programmable delay block in each receiver cell delays the receiver sampling point with respect to the transmitter clock. A distributed feedback laser array (Thorlabs PRO8) is used as the four channel optical source for the link. An arrayed waveguide grating multiplexes these different wavelengths of light from the laser array onto a single fibre. We place polarization controllers before the transmitter chip and before the receiver chip to ensure optimal coupling into the fundamental transverse electric (TE) mode of each waveguide. The optical switch is used to direct light from the transmitter to an erbium-doped fibre amplifier (EDFA) and an oscilloscope for each eye diagram; the switch then directs light back to the 3D integrated receiver for BER measurements without amplification. Optical losses in this link amount to 14.5 dB. These losses are from several sources: three chip-to-fibre interfaces at 3 dB each account for 9 dB, the modulation insertion loss is 2.5 dB, a modulation penalty accounts for 1.5 dB (this is the difference between the optical power of a ‘1’ bit and the average power), and an extra 1.5 dB of power is lost through the optical switch and fibre connectors throughout the link. \\

\textbf{Electronic circuit architectures.} Transmitted data starts as bits in the memory of each transmitter cell. Timing circuits running on a 5 GHz input clock (half transmitted data rate) generate memory read addresses and two-to-one multiplexer select signals. Circuits in the data path operate at a voltage supply of 1 V, except for the driver, which operates between 1 and 1.5 V. Extended Data Figure 3a shows the driver circuit. Inside of the design, high speed 1 V transistors in a cascode configuration prevents transistor junction breakdown from supply voltages exceeding 1 V. The main driver branch (M5-M8) has wide transistors to reduce the switching delay on modulator capacitance (C\textsubscript{microdisk} = 128 fF). A capacitor (C\textsubscript{coupling} = 183 fF) ensures a high switching speed while the auxiliary branch (M1-M4) holds the DC voltage level.

The receiver circuit senses a modulated photodiode current, amplifies it to digital levels at the supply rail voltage, and de-serializes the signal before writing it into internal memory. Extended Data Figure 3b shows the amplifier circuit, which uses an inverter-based trans-impedance amplifier (TIA) as an initial gain stage followed by an equalizer and inverters. A programmable current digital-to-analog converter (DAC) at the amplifier input supplies a current (IDAC) that cancels the DC offset of the photodiode current. The TIA stage has a high feedback resistance for a high gain (Rf = 18.6 k$\Omega$). This resistor equates to a lower input resistance (R\textsubscript{in} = 2.1 k$\Omega$) from the Miller Effect, however, it combines with the input capacitance (C\textsubscript{photodiode} = 17 fF, C\textsubscript{pad} = 10 fF) for a low frequency pole. As a remedy, an active inductor circuit in the subsequent equalizer stage cancels out the TIA pole (R\textsubscript{eq} = 3.1 k$\Omega$, C\textsubscript{eq} = 33.6 fF). After the equalizer, ensuing inverters ensure the output swings between 0 and 1 V. An isolated, 1 V power supply of the receiver amplifiers mitigates supply noise.

The TIA dominates the receiver amplifier energy consumption and its energy per bit is the static biasing power divided by the data rate. However, the TIA design introduces trade-offs between noise, bandwidth, and power. Equation 1 shows how the receiver signal-to-noise ratio, SNR, relates to receiver energy per bit (E/bit$_{ RX}$), input signal (I), input capacitance (C), and channel bandwidth (BW). Supplementary Note 3 provides a derivation of Equation 1. This relationship  sets a boundary on channel data rate scaling. With constant SNR and C, the design can expand BW with an increase in the input signal. In this context, the energy per bit remains constant, with the growing BW balancing out added laser power. This could imply an indefinite data rate scaling. However, a rise in BW necessitates wider TIA transistors, which subsequently contribute significantly to the input capacitance, C. This sequence results in a degradation of SNR at the receiver for high BWs, establishing a cap on the energy-efficient per-channel data rate. To achieve higher data rates without compromising energy, the focus should be on parallel data communication across multiple channels. Similar conclusions have been made in other studies, which advocate for parallel channels operating at moderate data rates \cite{miller2017attojoule,georgas2011addressing}.

\begin{equation}
SNR \sim \left ( \frac{I}{BW}\right )^2 \frac{E/bit_{RX}}{C^2}.
\end{equation}

\section*{Acknowledgements}

This work was supported by the U.S. Defense Advanced Research Projects Agency (DARPA) under PIPES Grant HR00111920014 and by the U.S. Advanced Research Projects Agency-Energy (ARPA-E) under ENLITENED Grant DE-AR000843. S.D acknowledges support by the National Science Foundation (NSF) GRFP under Grant DGE-1644869. We thank G. Keeler for leading the PIPES program, N. Abrams and M. Hattink for helpful discussions, and the engineering teams at AIM/SUNY Poly Photonics, Micross AIT, and CVI for their roles in the transceiver fabrication and assembly.

\section*{Author Contributions}

S.D. designed the photonic chip and developed the 3D bonding process. S.D. led transceiver testing and photonic device analysis with assistance from A.R., S.W., A.N., and V.G. A.R. compiled chip designs for the custom photonic wafer run. S.W., D.K., C.O., and A.M. designed the electronic chip and conducted bring-up tests of the electronic chip. S.D. and M.C. designed the printed circuit boards. R.P. gathered wafer-level micro-disk fabrication variation data. A.M. and K.B. supervised the project.

\section*{Competing Interests}

The authors declare no competing interests.

\section*{Data Availability}

The data that support the plots within this paper and other findings of this study are available from the corresponding author upon reasonable request.

\section*{Correspondence}

Correspondence and requests for materials should be addressed to K.B. (email: bergman@ee.columbia.edu).

\bibliographystyle{naturemag}
\bibliography{bibliography}

\renewcommand{\thefigure}{Extended Data 1}

\begin{figure*}[!h]
    \centering
    \includegraphics[scale=1]{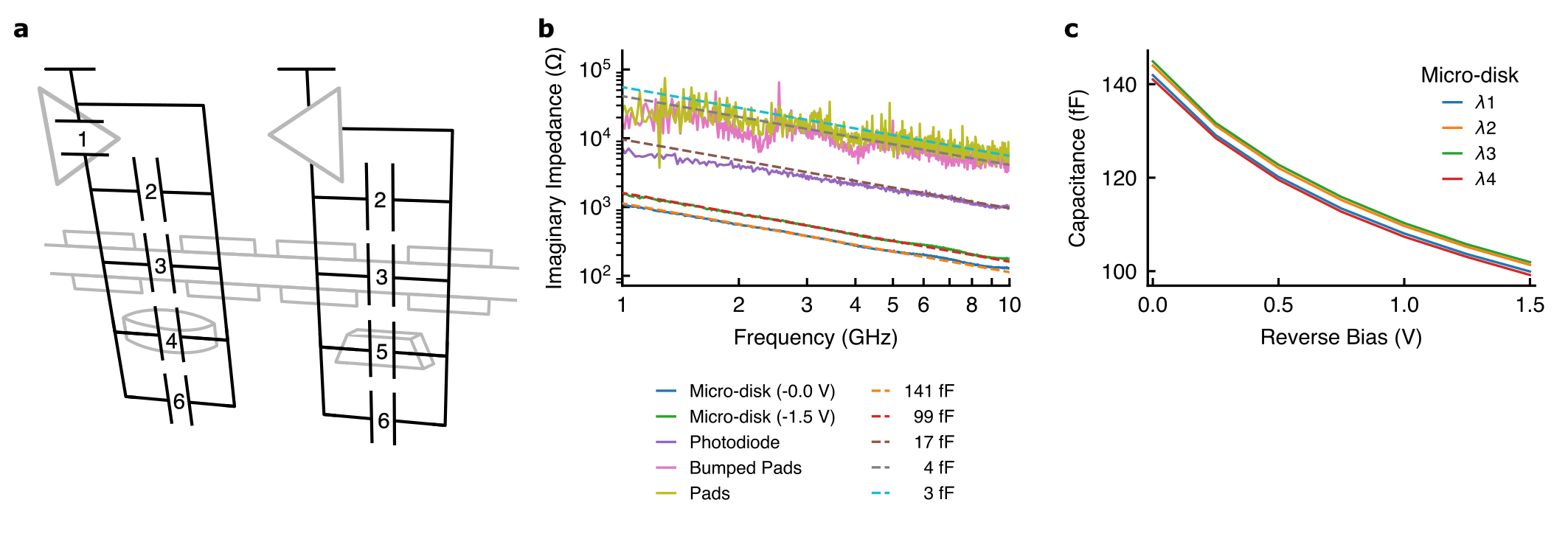}
    \caption{\textbf{Device capacitances.} \textbf{a,} Schematic of the transmitter and receiver capacitance sources; 1: electronic driver circuit, 2: electronic chip pads, 3: bump parasitics, 4: micro-disk PN junction, 5: photodiode PIN junction, 6: photonic chip pads. \textbf{b,} Measured imaginary impedances of devices and their fitted capacitor impedances. \textbf{c,} Measured micro-disk junction capacitance as a function of reverse bias voltage.}
    \label{fig:device_capacitance}
\end{figure*}

\renewcommand{\thefigure}{Extended Data 2}

\begin{figure*}[!h]
    \centering
    \includegraphics[scale=1]{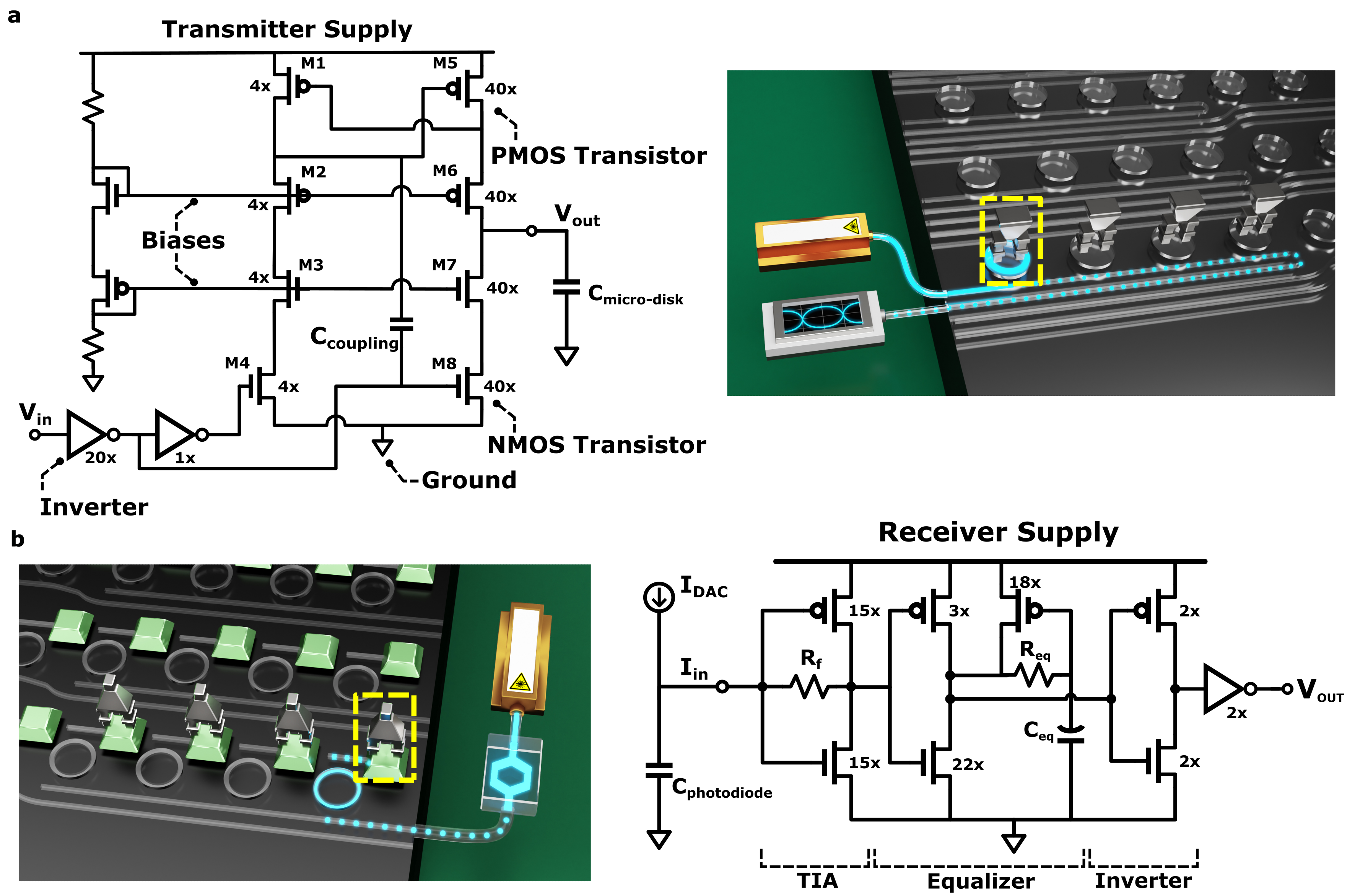}
    \caption{\textbf{Transmitter and receiver electronic circuit schematics.} \textbf{a,} Transmitter driver circuit schematic and driver highlighted in the transmitter illustration; data as V\textsubscript{in} drives the micro-disk as V\textsubscript{out}. \textbf{b,} Receiver circuit schematic and the receiver highlighted in the receiver illustration; amplification converts I\textsubscript{in} to V\textsubscript{out}. In the schematics, a multiplier labels each transistor width as a multiple of a ‘1x’ transistor. The ‘1x’ transistor has a width of 500 nm and length of 30 nm in the transmitter and 300 nm by 30 nm in the receiver.}
    \label{fig:comb_stability}
\end{figure*}

\end{document}

\preprint{APS/123-QED}

\title{Supplementary Information: 3D photonics for ultra-low energy, high bandwidth-density chip data links}% Force line breaks with \\

%\title{Massively scalable silicon photonic transmitter driven \\ by an integrated Kerr frequency comb}

\author{Stuart Daudlin\textsuperscript{1}, Anthony Rizzo\textsuperscript{1,2}, Sunwoo Lee\textsuperscript{3}, Devesh Khilwani\textsuperscript{3}, Christine Ou\textsuperscript{3}, Songli Wang\textsuperscript{1}, Asher Novick\textsuperscript{1}, Vignesh Gopal\textsuperscript{1}, Michael Cullen\textsuperscript{1}, Robert Parsons\textsuperscript{1}, Alyosha Molnar\textsuperscript{3}, and Keren Bergman\textsuperscript{1,*}}%

\address{\textsuperscript{1}Department of Electrical Engineering, Columbia University, New York, NY 10027 \\
\textsuperscript{2}Air Force Research Laboratory Information Directorate, Rome, NY 13441 \\
\textsuperscript{3} Department of Electrical and Computer Engineering, Cornell University, Ithaca, NY 14853}

% \date{\today}

\maketitle

\section*{Supplementary Note 1: Light Source Energy Consumption}

While a detailed analysis of the energy consumption of the electronic-photonic interface is provided in the main text, the energy consumption associated with the light source must also be considered to give a comprehensive evaluation of the total system energy. In this Supplementary Note, we provide an estimated laser energy consumption for two distinct classes of light sources: (i) distributed feedback laser arrays and (ii) integrated frequency comb sources. Furthermore, we show a detailed path to improving the link budget through experimentally validated low-loss chip-fiber couplers fabricated in the same process flow used for the photonic chips demonstrated in the main text. Through the reduced losses from these improved coupling interfaces, the required optical power from the laser sources are reduced and thus the associated energy consumption is projected to be substantially lower in future iterations of the system.  \\

\subsection*{Reduction of Chip-Fiber Interface Losses}

\begin{figure*}
    \includegraphics{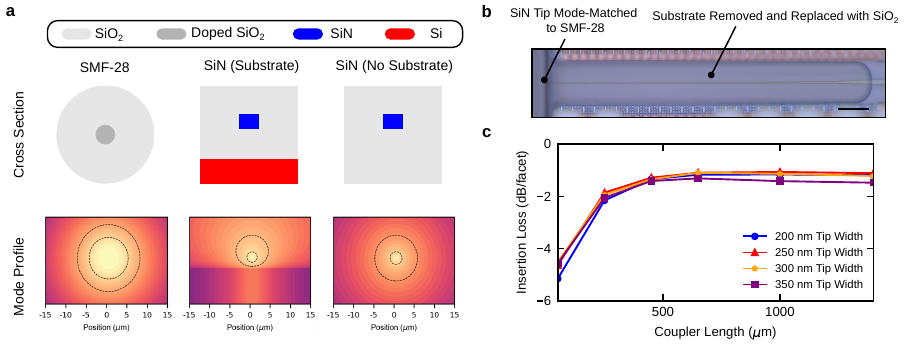}
    \caption{\textbf{Suspended edge couplers. a,} Cross-sectional views and simulated mode profiles for SMF-28 fiber, silicon nitride (SiN) edge coupler without substrate removal, and SiN edge coupler with selective substrate removal. Through removing the substrate, it is clear that the symmetry of the edge coupler modal profile more closely overlaps with that of SMF-28 fiber. \textbf{b,} Annotated microscope image of a fabricated suspended edge coupler with the removed substrate clearly visible. \textbf{c,} Experimentally measured test device results for various edge coupler tip widths as a function of taper length showing an optimal coupling loss of -1.1 dB/facet.} 
    \label{fig:fig_edge_couplers}
\end{figure*}

The total losses accrued across the full data communication link directly dictate the required optical power for each wavelength, and thus decreasing these losses directly results in reduced energy consumption of the optical source. In this demonstration, one of the main loss mechanisms is through each chip-fiber interface which occurs at three points in the link (source to transmitter, transmitter to fiber, and fiber to receiver). This loss is greatly improved relative to standard silicon photonic edge couplers by implementing suspended edge couplers which reduce the modal overlap between the inverse taper tip and the silicon substrate. Since the waveguide mode field diameter (MFD) is expanded using an inverse taper structure to match a standard SMF-28 fiber mode (MFD $\approx$ 10 $\mu$m) \cite{almeida2003nanotaper}, this large mode has substantial overlap with the silicon substrate at the chip facet since the edge coupler tip is only vertically separated from the substrate by 2 $\mu$m of buried oxide (BOX). This overlap distorts the symmetry of the modal shape, resulting in substantial mode-mismatch loss between the edge coupler and fiber.  However, this modal overlap with the substrate can be eliminated through selective substrate removal beneath the edge coupler, which is then replaced with silicon dioxide with an index of refraction matched to the rest of the oxide cladding. \\

\begin{figure}
    \includegraphics{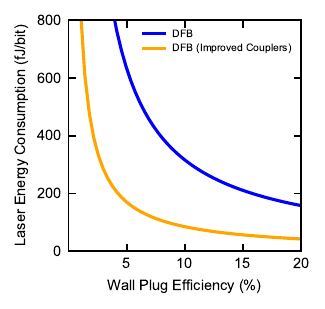}
    \caption{\textbf{DFB array energy consumption.} Energy consumption per bit for a DFB laser array as a function of laser WPE. At 18\% WPE \cite{koch2013integrated}, the array consumes 175 fJ/bit with the demonstrated edge couplers and 47 fJ/bit with improved edge couplers.}
    \label{fig:fig_laser_energy_dfb}
\end{figure}

Fig. \ref{fig:fig_edge_couplers}a shows the simulated modes for the fiber and edge coupler both for the standard design (including silicon substrate) and the suspended design (with the silicon substrate removed). From these simulations, it is clear that the suspended design yields better modal overlap between the fiber and edge coupler due to the improved symmetry. We implemented these suspended edge couplers in the foundry process used for the photonic chips in the demonstrated electronic-photonic transceiver (AIM Photonics). This process change was implemented at wafer-scale on 300 mm wafers. While the suspended edge couplers in the transceiver demonstrated a coupling loss of 3 dB per facet, this loss deviates from the simulated performance due to the absence of a deep trench etch on the bumped photonic wafers. Without this deep trench etch, the chips are mechanically diced without a smooth facet, thus adding additional loss. However, test devices in the same process including the deep trench etch exhibited experimentally measured coupling losses as low as 1.1 dB/facet with standard SMF-28 fiber and index matching fluid (Fig. \ref{fig:fig_edge_couplers}c). In future link implementations using these improved devices with both the selective substrate removal and deep trench etch with polished facet, the link budget can be improved by 5.7 dB (1.9 dB improvement at each fiber-chip interface) to yield 8.8 dB total link losses. With the experimentally measured receiver sensitivity of -19.5 dBm, this indicates that the optical power required from the laser at each wavelength can be reduced from -5 dBm to -10.7 dBm (0.085 mW).

\begin{figure}
    \includegraphics{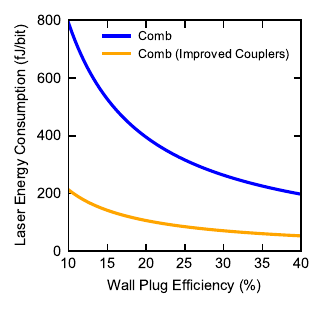}
    \caption{\textbf{Kerr comb energy consumption.} Energy consumption per bit for a Kerr comb source pumped by a DFB as a function of pump DFB WPE. At 35\% DFB WPE \cite{morrison2019high} and 40\% pump-to-comb conversion efficiency \cite{kim2019turn}, the comb consumes 225 fJ/bit with the demonstrated edge couplers and 60 fJ/bit with improved edge couplers.}
    \label{fig:fig_laser_energy_comb}
\end{figure}

\subsection*{Distributed Feedback Laser Array Energy Consumption}

Distributed feedback (DFB) laser arrays represent a mature multi-wavelength light source for wavelength division multiplexing (WDM) and have been shown in numerous demonstrations for applications in optical communications \cite{wade2021error, kumar2022demonstration, li2014eight}. Since DFB lasers tend to exhibit peak wall-plug efficiency (WPE) at higher optical output powers, the best achievable WPE rises from around 18\% at 10 mW optical power \cite{koch2013integrated} to 35\% at 250 mW optical power \cite{morrison2019high}. To leverage this inherent property of the laser operation for optimal energy allotment, we assume an array of high power DFB lasers at each WDM wavelength which are multiplexed together and then split into various ports where each wavelength in each port has just enough power to overcome the link losses \cite{buckley2018wdm}. However, splitting 250 mW of optical power into streams of -5 dBm quickly becomes impractical, and thus we assume that in the DFB case the ideal laser operating condition is at 10 mW with 18\% WPE. For simplicity in the analysis, we treat each wavelength in each port after splitting as its own independent source (``effective laser''). Given the experimentally measured optical link losses of 14.5 dB for the amplifier-free chip-to-chip link demonstrated in the main text and receiver sensitivity of -19.5 dBm, this yields an optical power requirement of -5 dBm for each wavelength with standard edge couplers and -10.7 dBm with improved edge couplers. Assuming a WPE of 18\% at -5 dBm (0.316 mW) optical power after splitting, each ``effective laser'' consumes 1.76 mW of electrical power. At a data rate of 10 Gb/s per ``effective laser'', this yields a total energy consumption of 175 fJ/bit. For the case of improved edge couplers, this energy consumption drops to 47 fJ/bit. Since the WPE of DFBs can vary significantly between designs, the energy consumption per bit is provided as a function of WPE in Fig. \ref{fig:fig_laser_energy_dfb}. However, through moving to integrated Kerr frequency combs which rely on higher pump power for seamless conversion into new frequency channels, the higher WPE regime of high-power DFB lasers can be accessed without excessive splitting penalties.

\subsection*{Integrated Comb Source Energy Consumption}

\begin{figure}
    \includegraphics{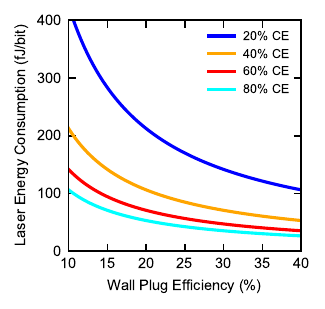}
    \caption{\textbf{Effect of Kerr comb conversion efficiency (CE) on energy consumption.} Energy consumption per bit for a Kerr comb source pumped by a DFB as a function of pump DFB WPE for various pump-to-comb conversion efficiencies. At 35\% DFB WPE \cite{morrison2019high} and 80\% pump-to-comb conversion efficiency \cite{zang2022near}, the comb consumes 113 fJ/bit with the demonstrated edge couplers and 30 fJ/bit with improved edge couplers. All curves are plotted assuming improved edge couplers.}
    \label{fig:fig_laser_conversion_comb}
\end{figure}

While DFB arrays are a mature solution for WDM systems, integrated comb sources are also highly appealing due to their ability to provide many wavelength channels from a single device \cite{gaeta2019photonic, chang2022integrated, rizzo2023massively, rizzo2022petabit}. Recent demonstrations have shown that integrated Kerr frequency combs can be coherently combined to boost the power-per-line while also permitting spectral shaping to flatten the comb spectrum \cite{kim2023coherent, kim2021synchronization}, making their deployment as WDM sources in future photonic interconnects a realistic prospect. In particular, nonsolitonic Kerr frequency combs in the normal group velocity dispersion (GVD) regime exhibit much higher pump-to-comb conversion efficiencies compared to soliton Kerr combs in the anomalous GVD regime \cite{kim2019turn, jang2021conversion}. While standard coupled-resonator designs for nonsolitonic Kerr combs can achieve conversion efficiencies (CE) as high as 41\% \cite{kim2019turn}, recent demonstrations of advanced designs have shown pump-to-comb CE approaching unity (86\%) \cite{zang2022near}. The comb source energy consumption for both edge coupler cases is shown in Fig. \ref{fig:fig_laser_energy_comb} as a function of pump laser WPE. Fig. \ref{fig:fig_laser_conversion_comb} shows the energy per bit as a function of pump DFB WPE for various pump-to-comb CEs. Using a high power DFB pump with 35\% WPE and a pump-to-comb CE of 40\%, the energy consumption of the comb source is 60 fJ/bit (assuming improved edge couplers). However, using an improved CE of 80\% as demonstrated in ref. \cite{zang2022near}, this energy consumption improves by a factor of 2 to 30 fJ/bit.

\section*{Supplementary Note 2: Thermal Tuning Energy Consumption}

\begin{figure}
    \includegraphics{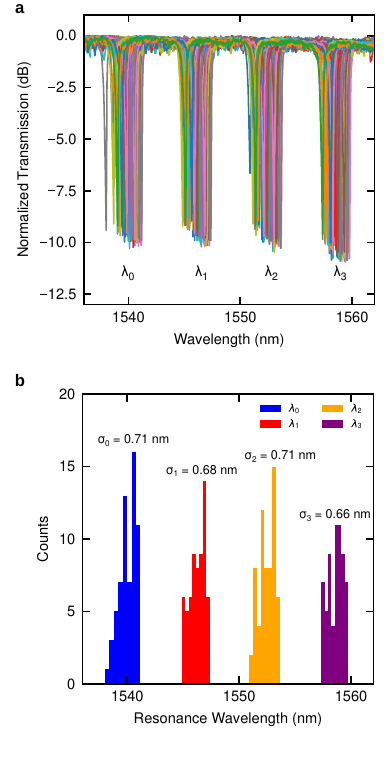}
    \caption{\textbf{Microdisk modulator resonance variation. a,} Measured spectrum for resonator buses across 62 reticles on a 300 mm wafer. Each bus contains four resonators, each targeted at a different nominal wavelength ($\lambda_0$, $\lambda_1$, $\lambda_2$, and $\lambda_3$). \textbf{b,} Histogram of resonance wavelength for each nominal device across all 62 reticles, with the standard deviation labeled for each target wavelength. The mean of the standard deviation across all devices is $\sigma_{avg} = 0.69$ nm.}
    \label{fig:fig_modulator_resonance_variation}
\end{figure}

\begin{figure}
    \includegraphics{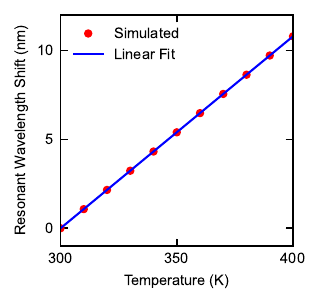}
    \caption{\textbf{Temperature sensitivity of microdisk modulators.} Simulated temperature dependence for a microdisk modulator with a free spectral range (FSR) of approximately 24 nm near $\lambda = 1,550$ nm. The slope of the linear fit is 108 pm/K.}
    \label{fig:fig_temperature_dependence}
\end{figure}

Since the laser wavelengths are assumed to be fixed under realistic field conditions in a deployed system, the resonators must be tuned to align with the laser wavelength grid. The two contributions to the total required resonator tuning are deviations due to fabrication variations (static) and deviations due to temperature fluctuations (dynamic). The magnitude of these variations are shown in Fig. \ref{fig:fig_temperature_dependence} using realistic values for process variations and thermal swings. The simulations assume a thermo-optic coefficient of $1.8 \times 10^{-4}$ K\textsuperscript{-1} for intrinsic silicon near room temperature for $\lambda =$ 1,550 nm \cite{komma2012thermo}. \\

The resonator free spectral range (FSR) was measured to be approximately 24 nm near a wavelength of 1,550 nm, and a full FSR corresponds to a $2 \pi$ phase shift. Therefore, the slope of the curve shown in Fig. \ref{fig:fig_temperature_dependence} is equivalent to 0.03 radians/K. While the electrical power required to yield a $\pi$ phase shift ($P_{\pi}$) is approximately 30 mW for standard microdisk modulator integrated heater designs, we have recently demonstrated improved designs with a selective substrate undercut which exhibit $P_{\pi} = 6.9$ mW \cite{rizzo2023ultra}. \\

While the thermal fluctuations of the photonic chip result in resonator drift, the integrated heaters in each resonator are additionally used to trim the resonance wavelength to compensate for fabrication variations. Using experimentally measured wafer-scale data for resonators across a full 300 mm wafer, we quantify the magnitude of the resonance variations to be $\sigma = 690$ pm per device (Fig. \ref{fig:fig_modulator_resonance_variation}). We use this value together with the experimentally measured 6.9 mW $P_{\pi}$ to quantify the energy consumed by each resonator due to fabrication variations. The device tuning efficiency is 0.45 radians/mW, or 1.74 nm/mW. Thus, to correct for a one $\sigma$ resonance variation of 690 pm, we require 0.4 mW per resonator. At a data rate of 10 Gb/s, this corresponds to 40 fJ/bit energy consumption required to correct for fabrication variations. If we consider $3\sigma$ variations, the energy consumption rises to 120 fJ/bit. \\

\begin{figure}
    \includegraphics[scale=1]{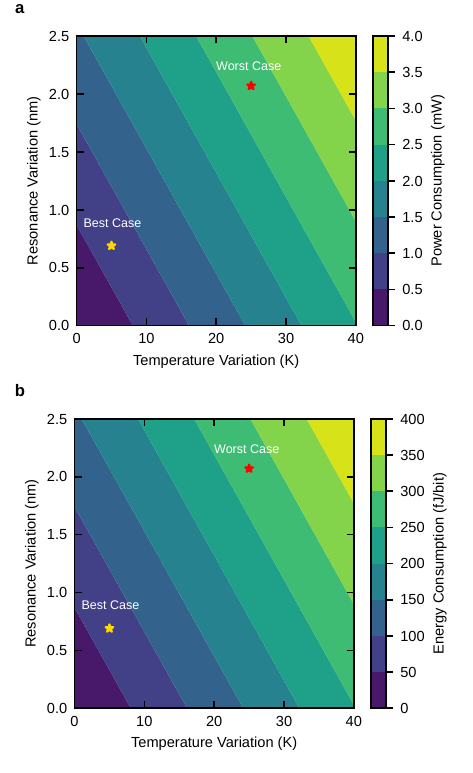}
    \caption{\textbf{Power and energy consumption of microdisk modulators. a,} Calculated power of each resonant device in milliwatts as a function of resonance variation due to fabrication and temperature variation. \textbf{b,} Calculated energy consumption per bit showing the `best case' value of 71 fJ/bit and `worst case' value of 274 fJ/bit.}
    \label{fig:fig_resonator_energy}
\end{figure}

In the energy consumption analysis, we consider corners based on `best case' and `worst case' values for both the fabrication variations and thermal fluctuations. We assume that under a normal, quasi-constant load, the thermal environment for the photonic devices will fluctuate on the order of $\Delta$T = 10 K. If we assume the temperature offset from baseline at each time slice follows a normal distribution defined over the range $\Delta T = 0$ K to $\Delta T = 10$ K, the mean temperature offset is $\Delta T = 5$ K. Under more dynamic thermal loads, we assume that the photonic devices can fluctuate up to 50 K and then following the same logic, the average offset is $\Delta T = 25$ K. For the `best case' fabrication variations, we assume one $\sigma$ values, whereas for the `worst case' we assume 3$\sigma$ values. These corners are summarized in the plot shown in Fig. \ref{fig:fig_resonator_energy},  with the `best case' energy consumption calculated at 71 fJ/bit and `worst case' energy consumption calculated at 274 fJ/bit.

\clearpage

\section*{Supplementary Note 3: Transimpedance Amplifier Energy Consumption}

We explore the trade-offs between noise, bandwidth, and power of the receiver TIA by starting with the input-referred thermal noise current per Hz of the TIA, 
\begin{equation}
\frac{\overline{i^2_{n,in}}}{Hz} = \frac{4kT}{R_f} + \frac{4kT}{2g_m}\gamma C^2(2\pi f)^2
\end{equation}
where $k$ is the Boltzmann constant, $T$ is temperature, $\gamma$ is the excess noise factor of the transistors, $f$ is frequency, $R_f$ is the TIA feedback resistance, $g_m$ is the transistor transconductance, and $C$ is the TIA input capacitance. The second term dominates the noise; integrating this term over frequency from zero to the channel bandwidth ($f$ = $BW$) and removing constants, 
\begin{equation}
\overline{i^2_{n,in}} \sim \frac{1}{g_m}C^2BW^3
\end{equation}
where $\overline{i^2_{n,in}}$ is the total input-referred noise current. The signal-to-noise ratio, $SNR$, is the squared input signal current, $I^2$, divided by the total input noise,
\begin{equation}
SNR \sim I^2\frac{g_m}{C^2BW^3}.
\end{equation}
Seeing that $g_m$ is proportional to  receiver static biasing power and BW is proportional to bits/second,
\begin{equation}
SNR \sim \left ( \frac{I}{BW}\right )^2 \frac{E/bit_{RX}}{C^2}.
\end{equation}

% We take the square root of $\overline{i^2_{n,in}}$ and label it $N$, the root-mean-squared input-referred current noise. Then, removing constants,
% \begin{equation}
% N \sim \frac{1}{g_m^{1/2}}\cdot C \cdot BW^{3/2}
% \end{equation}
% and seeing that $g_m$ is proportional to  receiver static biasing power and BW is proportional to bits/second,
% \begin{equation}
% N \sim \frac{1}{(E/bit_{ RX})^{1/2}}\cdot C\cdot BW
% \end{equation}
% or,
% \begin{equation}\label{eq:rx}
% E/bit_{ RX} \sim (C\cdot \frac{1}{N}\cdot BW)^2
% \end{equation}

% \section*{Supplementary Note 3: Electronic Integrated Circuit Block Diagram and Layout}

% \section*{Supplementary Note 4: MCM-to-MCM Demonstration Experimental Setup}

\bibliographystyle{naturemag}
\bibliography{bibliography}